\documentclass{jfm}
\usepackage{graphicx}
\usepackage{epstopdf, epsfig}
\usepackage{amsmath}
\usepackage{hyperref}
\usepackage{multirow}
\usepackage[dvipsnames]{xcolor}

\hypersetup{hidelinks}

\usepackage[amssymb]{SIunits}
\usepackage{color}
\usepackage[normalem]{ulem}

\DeclareMathOperator{\erf}{erf}

\makeatletter
\newcommand\thefontsize{The current font size is: \f@size pt}
\makeatother

\shorttitle{Melting of olive oil in immiscible surroundings: experiments and theory}
\shortauthor{P. Waasdorp,  A. van den Bogaard, L. van Wijngaarden, S. G. Huisman}

\title{Melting of olive oil in immiscible surroundings: experiments and theory}

\author{Pim Waasdorp, Aron van den Bogaard,\\Leen van Wijngaarden, \and Sander G. Huisman\corresp{\email{s.g.huisman@utwente.nl}}}

\affiliation{Physics of Fluids Group and Max-Planck Center for Complex Fluid Dynamics, Faculty of Science and Technology, J.M. Burgers Center for Fluid Dynamics, University of Twente\\7500 AE Enschede, The Netherlands}

\begin{document}

\maketitle

\begin{abstract}
We report on the melting dynamics of frozen olive oil in quiescent water for Rayleigh numbers up to $10^9$. The density difference results in an upward buoyancy-driven flow of liquid oil forming a thin film around the frozen oil. We experimentally investigate flat, cylindrical, and spherical shapes and we derive theoretical expressions for the local film thickness, velocity, and the local melt rate for these three canonical geometries. Our theoretical models predict the correct order of magnitude and the correct scaling as compared to our experimental findings.\\

\hfill \break
\end{abstract}

\section{Introduction}
Understanding the complicated dynamics of phase change is relevant to predict and control many natural and industrial processes. Melting and dissolution are examples of the classical Stefan problem, where the boundary is defined by the phase of the material and the evolution of the boundary follows from the material undergoing phase change. Common examples include the freezing of water to make ice cubes, novel developments in phase change materials, where the latent heat of fusion is used as a temporary energy storage \citep{Dhaidan2015MeltingReview}, and the melting of ice around Earth's North and South Poles \citep{Holland2006FutureIce,Feltham2008,Cenedese2023}.

During melting, the cold melt generally flows along the body, giving rise to non-uniform melting, and therefore changing the shape of the object, which then feeds back on the flow. This shape change or self-sculpting process of objects subject to melting, erosion, or dissolution has been a topic of recent interest. The evolution of eroding clay balls and cylinders in a flow have been studied by \cite{Ristroph2012}. More recently, more studies have been done with quiescent surroundings, as in \cite{Cohen2016,DaviesWykes2018,Pegler2020,Cohen2020}, where they studied the pattern formation due to natural convection and dissolution of hard candy and salt, submerged in water. Pattern formation was also studied by \cite{Guerin2020} who, using experiments, reveal the dynamics of karst geomorphology and rillenkarren formations. Further insights into the emergence of rock formations due to dissolution are provided by \cite{DaviesWykes2018} and \cite{Huang2020}, who emphasise the importance of the directionality of the shaping process. In \cite{Huang2022} a class of exact solutions is given for the shape of the pinnacles, which show that the tip curvature is large, but finite. Recently there have been direct numerical simulations by \cite{Yang2023IceSalinity,Yang2023MorphologyBelow}, who use the phase-field method to study the morphology of melting ice in a Rayleigh--B\'{e}nard geometry and stratification of salt concentration around a melting cylinder.

Hitherto, most studies have focused on the melting of miscible fluids, i.e. a frozen object submerged in the same substance in liquid phase, or a similar miscible liquid or solution (e.g.~melting of ice in salty water). The case of immiscible melting has received little attention. Immiscible melting can be achieved in multiple ways: either we have organic compounds like oils and waxes and combine that with water, or we have metals (e.g.~gallium) inside water or oils. The most experimentally-accessible option is to use an oil with a freezing point around $\unit{0}{\celsius}$ and water. Motivated in the context of nuclear meltdown accidents in nuclear reactors involving molten core material, \cite{taghavi1979thermal} looked at the case of melting of an immiscible liquid where the frozen oil is below warmer water. In the melting of a horizontal wall the process is governed by the Rayleigh--Taylor instability \citep{gennes2004capillarity}, as the melting material (oil) is positively buoyant in the water. The interface in their case is therefore heavily undulated by the pinching-off droplets. Nevertheless, they have found a scaling of $\text{Nu} \propto \text{Ra}^{1/4}$, despite all the intricacies associated with the pinching-off droplets and the non-uniform liquid oil layer. These findings were followed up by \cite{farhadieh1982downward} where they looked at a variety of oily or waxy substances melting in a variety of watery solutions. They observe from their experimental measurements that their data is bounded by two different $\text{Nu} \propto \text{Ra}^{1/4}$ scaling laws, but that their data follows more closely a $\text{Nu} \propto \text{Ra}^{1/5}$ scaling set forth by \cite{gerstmann1967laminar} in terms of absolute agreement, though they did not make any definitive conclusion on the scaling exponent due to the scatter in their experimental findings. In this work we study the melting process of frozen olive oil in water around room temperature, as this is most accessible experimentally. We will, however, focus on configurations where the Rayleigh--Taylor instability only plays a minor role.

For thermal convection problems with phase-change the three dimensionless control parameters are the Rayleigh ($\text{Ra}_o = \frac{g(\rho_w-\rho_o)L^3}{\nu_o\alpha_o\rho_o}$), Stefan ($\text{Ste} = \frac{c_p\Delta T}{\mathcal{L}}$), and Prandtl ($\text{Pr}=\nu_o/\alpha_o$) numbers, where $g$ is gravity, $\rho_w$ is the density of the water, $\rho_o$ is the density of the olive oil, $L$ is a relevant length scale, $\nu$ is the kinematic viscosity, $\alpha$ is the thermal diffusivity, $c_p$ is the specific heat, $\Delta T = T_\infty - T_o$ is the temperature difference between the ambient and the melting temperature, and $\mathcal{L}$ is the latent heat of fusion. We have used $o$ and $w$ subscripts to denote quantities related to oil and water, respectively. Here, the Rayleigh number is the time scale associated to thermal transport due to diffusion, as compared to the time scale of thermal transport due to convection. Whereas in classical Rayleigh--B\'enard convection buoyancy is created by (generally) small density changes due to temperature changes ($\beta \Delta T$), in our case a large density difference is immediately created due to the different substances $\left((\rho_w-\rho_o)/\rho_o\right)$. A high Rayleigh number means intense thermal driving of the system. The Stefan number describes the ratio of specific heat versus the latent heat of fusion, where the latent heat is the heat needed or released by a phase change. A higher Stefan number means that phase change happens faster. The Prandtl number is a material property describing the ratio of the momentum diffusivity to the thermal diffusivity, and determines whether the thermal boundary layer is embedded in the momentum boundary layer or vice versa.

The present work has the following structure: in section \ref{sec:expsetup} we describe the experimental setup. In section \ref{sec:results}, results for the melt rate of frozen olive oil are shown for three different geometries: a vertical wall, a cylinder, and a ball. For the cylinder we show two different initial Rayleigh numbers (initial sizes). The obtained local melt rates are compared with theoretical models that are derived in detail in section \ref{sec:theory}. The effect of the assumption of constant viscosity is discussed, and a correction for the variation of the viscosity with temperature is derived. Lastly, we discuss our findings in detail in section \ref{sec:discussion} and finish with our conclusions in section \ref{sec:conclusion}.

\section{Experimental setup} \label{sec:expsetup}
\begin{figure}
    \centering
    \includegraphics[width=.9\columnwidth]{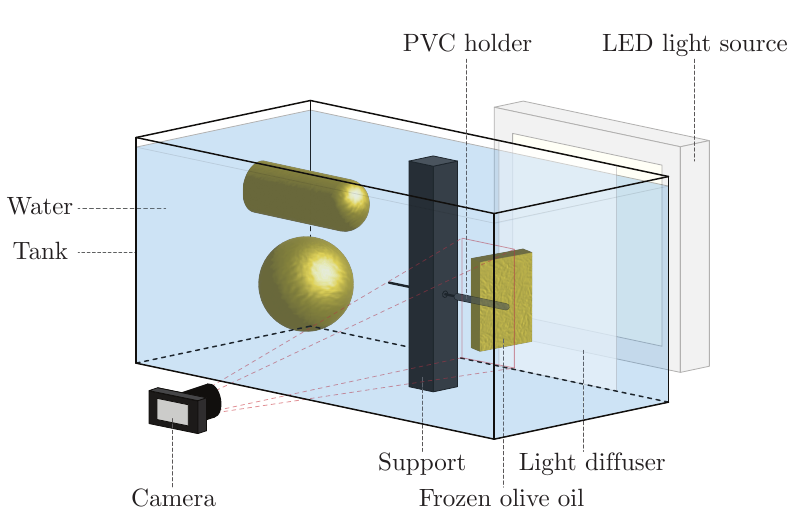}
    \caption{Schematic of the experimental setup. The frozen olive oil object is submerged in quiescent water. The dimensions of the glass tank are $\unit{400}{\milli\meter}\times \unit{500}{\milli\meter}\times \unit{800}{\milli\meter}$. A PVC holder is incorporated in the frozen oil during freezing, and connects the frozen oil to a support. A white LED light source is used with a light diffuser to create a uniform background illumination. Three canonical geometries are shown: vertical wall (photographed), horizontal cylinder, and ball. A camera periodically photographs the melting objects from the side (vertical wall) or the front (cylinder and ball).}
    \label{fig:expsetup}
\end{figure}

\begin{figure}
    \centering
    \includegraphics[width=1\columnwidth]{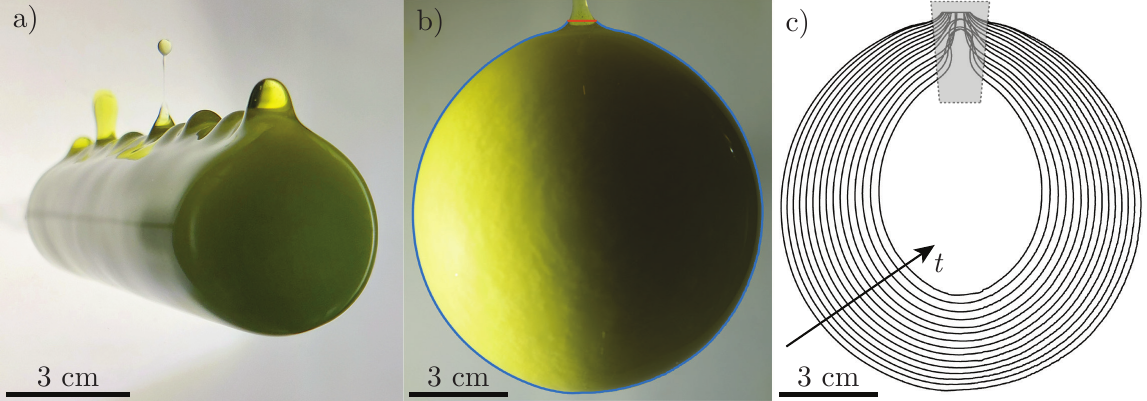}
    \caption{a) Melting of a horizontal cylinder in ambient water showing several stages of the droplet pinch-off at the top where the melted olive oil collects. Note that we consider the melting cylinder only in 2D cross section in this work. The distance between the pinching-off droplets is $\unit{22.5}{\milli\meter}\unit{\pm 2}{\milli\meter}$. b) Example image from a melting ball experiment with the contour that is obtained from image analysis overlaid on the original image. Note that the region where the liquid olive oil melt collects and periodically pinches off from the ball is ignored. c) Contours of a melting ball for $t=\unit{200}{\second}$ to \unit{3200}{\second}, with intervals of \unit{200}{\second}. Note that the tracking on the apex of the ball is hindered by the collecting and periodically pinching-off of liquid olive oil melt. Therefore, data is ignored in the grey shaded region.}
    \label{fig:contourcomp}
\end{figure}

The schematic in figure \ref{fig:expsetup} shows the experimental setup we use to study the melting of frozen olive oil. We choose olive oil as our working fluid since the freezing/melting temperature is reasonable to achieve in the lab ($T_o = \unit{-8}{\degreecelsius}$), the resulting contact temperature is above the freezing temperature for the ambient water, and it is readily available. We use a rectangular glass tank of $\unit{400}{\milli\meter}\times \unit{500}{\milli\meter}\times \unit{800}{\milli\meter}$, filled with water. During the melting process, the water is quiescent and kept at a constant room temperature $T_\infty = \unit{20}{\celsius}$. We only investigate here for ambient temperatures equal to this room temperature, such as to avoid any spurious flows created by natural convection caused by heat transfer from the surroundings to the bath. A small PVC holder (poor thermal conductivity) is included in the frozen olive oil during the freezing process and is attached to a support that is submerged in water.  The olive oil is cooled down to a temperature of $T_i = \unit{-14}{\celsius}$. Further material properties of both substances can be found in table \ref{tab:liqprop}, including the surface tension. Since $\rho_{\text{oil}} < \rho_{\text{water}}$ the melted olive oil will rise and collect at the apex of the objects where it periodically pinches off. There the surface tension is relevant for the pinch-off dynamics. We do not study this effect here and will ignore the surface tension, which we elaborate on in the discussion section. For studies of droplet deformation and breakup in viscous fluids we refer to e.g.\ \cite{Stone1994,Nagel1999,Nagel2001}. We study three different canonical geometries: a vertical wall, a horizontal cylinder, and a ball, see figure \ref{fig:expsetup}. The melting process is recorded through interval imaging. For this, a DSLR camera (Nikon D850) with a \unit{100}{\milli\meter} macro objective (Zeiss Makro Planar T* 2/100) is used, resulting in a resolution of \unit{30}{\micro\meter\per px}. An LED light source and light diffuser are used to create a uniformly-lit background. The original videos can be found in the supplemental materials. The images are binarized after which we find the contour, area, and, for the cylinders and balls, the centroid of the object, see figures \ref{fig:contourcomp}b and \ref{fig:contourcomp}c. From the evolution of the contour we find the local melt rate.

\begin{table}
\begin{center}
\begin{tabular}{ccccccccccc}
          & $T_m$ & $\rho$ & $\mu$ & $\alpha$ & $\lambda$ & $c_p$ & $\mathcal{L}$  & $\sigma$ & Pr & Ste \\
          & $\left[\celsius \pm \kelvin \right]$ & $\left[\frac{\kilo\gram}{\meter^3}\right]$ & $\left[\frac{\gram}{\meter \cdot \second}\right]$ & $\left[\frac{\milli\meter^2}{\second}\right]$ & $\left[\frac{\watt}{\kelvin \cdot \meter}\right]$ & $\left[\frac{\joule}{\kilo\gram \cdot \kelvin}\right]$ & $\left[\frac{\kilo\joule}{\kilo\gram}\right]$ & $\left[\frac{\milli\newton}{\meter}\right]$ & $[-]$ & $[-]$\\ \hline
Water     & $0$   & $999.7$  &  $1.304$   &     $0.138$  &  $0.58$    & $4192$ & $334$   & $23^*$   & $9.45$      & $-$ \\ 
          &       & $\pm0.1$ & $\pm0.001$ &  $\pm 0.001$ & $\pm0.01$  & $\pm1$ & $\pm 1$ & $\pm 1$  & $\pm 0.07$  &      \\ 
          \hline \\
Olive oil & $-8$       & $870$     & $170$    & $0.0796$     & $0.166$     & $1970$   & $267$   & $23^*$  & $2455$    & $0.207$\\ 
          &  $\pm 0.5$ & $\pm 10 $ & $\pm 5$ & $\pm 0.0038$ & $\pm 0.001$ & $\pm 10$  & $\pm 1$ & $\pm 1$ & $\pm 141$ & $\pm 0.004$\\
          \hline
\end{tabular}\caption{Material properties of water and olive oil. Properties for water are taken at $T = 10\degree$C, following \cite{Bejan1993}. Thermal properties for olive oil are taken from \cite{Valdez2006,Turgut2009}. Values are taken for a constant temperature of $\unit{3}{\celsius}$; the mean temperature in the olive oil melt layer. The value for the latent heat of fusion is taken from \cite{LatentHeatTable}. Comparing with values for other vegetable oils from \cite{Gudheim1944} we find that they are of comparable magnitude. For quantities given without error estimates in their source we have assumed the last digit to be indicating the precision. Note that the thermal properties of olive oil are not entirely self-consistent:  $\alpha \rho c_p/\lambda = 0.82 \pm 0.04$. This is due to the fact that not many different sources are available, and measurements were done on different olive oils, possibly with different compositions. Wherever one of these quantities is needed, we opt for the `most direct' quantity.  $^*$The surface tension $\sigma$ was measured using the pendant drop method for olive oil submerged in water at a temperature of $\unit{20}{\celsius}$.}\label{tab:liqprop}
\color{black}
\end{center}
\end{table}

The image processing is applied to all images that are taken during an experiment, typically with an interval time of $\Delta t = \unit{20}{\second}$. In figure \ref{fig:contourcomp}c contours from a single experiment are shown at different times. Such an image shows a qualitative description of the melting process of a ball. The contours at the top are more closely spaced, whereas contours at the bottom are more distant, revealing that the melt rate at the bottom of the ball is higher than the melt rate at the top. 

\section{Results} \label{sec:results}
Here we will look at the melt rates obtained from the experiments. We will then compare these to analytical expressions---derived in the next section---and discuss the applicability of the theory for the vertical wall, for the two horizontal cylinders, and lastly for a ball. 

\subsection{Vertical wall}
\begin{figure}
    \centering
    \includegraphics[width=0.8\columnwidth]{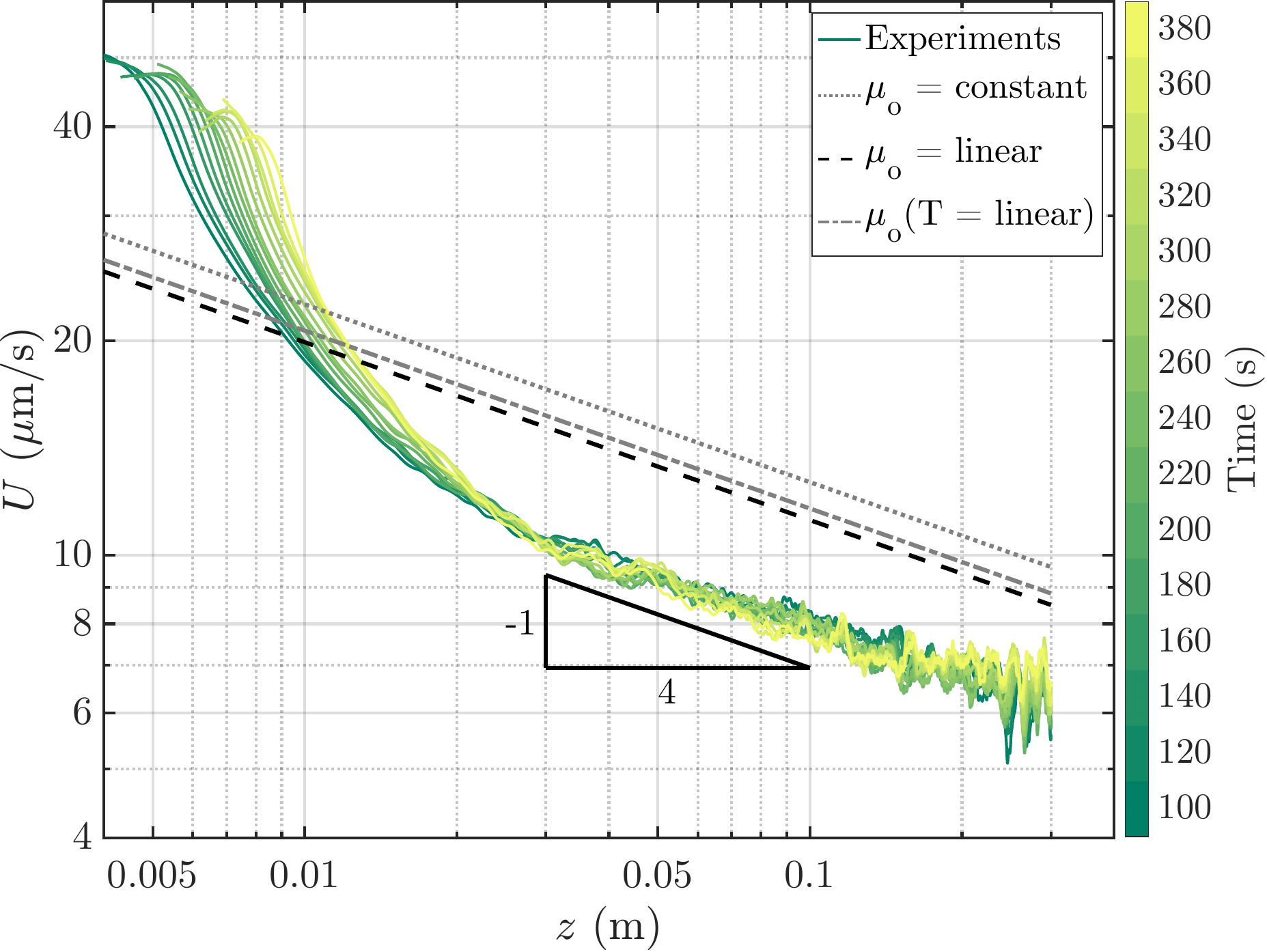}
    \caption{Horizontal melt-rate as a function of the height for the case of a vertical wall for $\text{Ra}=\mathcal{O}(10^9)$. We show three theoretical approximations, that are derived in a later section. Here the theoretical approximation with constant viscosity $\mu_o$ is shown as a grey dotted line, the black dashed line is assuming a linear viscosity profile in the melt layer, and the gray dash-dotted line is assuming variable viscosity resulting from a linear temperature profiles in the melt layer. The intermediate part of the profile shows a scaling of $-1/4$ with the height. Profiles for times from \unit{100}{\second} until \unit{380}{\second} from the start of the melting process are shown. At later times the vertical wall is no longer an appropriate approximation. The video of this experiment can be found in the supplemental materials.}
    \label{fig:wallmumodel}
\end{figure}

We first look at the case of the melting of a block of olive oil of \unit{30}{\centi\meter} height, see figure \ref{fig:theorywall} for the definitions of the axes and other quantities. We calculate the horizontal local melting rate from the evolving contours, see figure \ref{fig:wallmumodel}. The profiles in the early stages of the melting process, where the shape, despite slightly changing over time, can be regarded as vertical. At later times, the profile of the initially rectangular block has sculpted itself away from its rectangular shape. While the total process of melting takes about \unit{40}{\minute}, here we just show melt rates obtained during the first \unit{5}{\minute}. In the lower regions of the vertical wall ($z < \unit{2}{\centi\meter}$), finite-size effects of the object are observed (the bottom corner is rounded over time). We do not show the upper edge region of the melting wall, since the results are heavily influenced by the accumulation and detaching of oil droplets. Away from the top and bottom corners it can be seen that the melt rates are remarkably constant over time, in both scaling ($U\propto z^{-1/4}$) and magnitude. The grey dotted line shows the theoretical model with constant viscosity in the melt layer, whereas the black dashed line shows the theoretical model where the viscosity $\mu(T)$ varies in the melt layer (see section \ref{sec:varvisco}). We find that our analytical expression predicts the correct order of magnitude and the correct scaling in this region. Note that our model does not contain any fitting parameters, and even though several approximations have been made, there is order of magnitude and scaling agreement between the model and the experiments. The inclusion of the variable viscosity lowers the prediction of the melting rate and therefore improves the prediction. Henceforth, we only show models with temperature-dependent variable viscosity in the melting layer.

\begin{figure}
    \centering
    \includegraphics[width=1\columnwidth]{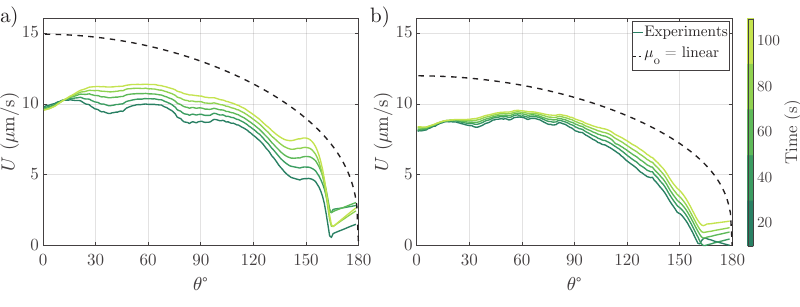}
    \caption{a) Cross-sectional melt-rate as a function of the angle $\theta$ (polar angle starting from the bottom, see figure~\ref{fig:theorycylinder}) for a small cylinder ($R_0 = \unit{25}{\milli\meter}$) $\text{Ra}=\mathcal{O}(10^7)$. The black dashed line shows the theoretical model for the cylindrical geometry. b) Large cylinder ($R_0 = \unit{60}{\milli\meter}$) $\text{Ra}=\mathcal{O}(10^8)$. The video of these experiments can be found in the supplemental materials.}
    \label{fig:cylindermeltrates}
\end{figure}

\subsection{Cylinder}
We perform experiments for horizontal cylinders with initial radii \unit{25}{\milli\meter} and \unit{60}{\milli\meter} corresponding to Rayleigh numbers $\text{Ra} \approx 10^7$ and $\text{Ra} \approx 10^8$, see figure \ref{fig:theorycylinder} for the definition of the axes and other measurements. Figure \ref{fig:cylindermeltrates} shows the radial melt rates $U(\theta) = dr(\theta)/dt$ for the small and large cylinder, as a function of the polar angle $\theta$, where $\theta = \unit{0}{\degree}$ is at the bottom of the cylinder. For the analysis we restrict ourselves to the early stages of melting, when the assumption of a circular cross--section is still valid, since this is one of the key geometrical assumptions in the theoretical model. The theoretical model, without any fitting parameter, for the cylindrical geometry is included as a black dashed line. Here the temperature-dependent viscosity is included in our model. A reasonable agreement is found between our experiments and our model for both cylinders in terms of order of magnitude and shape. There are some notable differences between the model and the experimental observations. The melting process of the cylindrical shape has a maximum melt rate along the surface at an angle of $\theta \approx 60\degree$ from the bottom, whereas the theory predict a monotonic decrease with increasing $\theta$, such that the predicted maximum melt rate is at the bottom. For $\theta \geq \unit{160}{\degree}$ the theory and experiments do not match due to the collection of melt at the top before pinching off \citep{shi1994cascade} at the top of the cylinder and rising to the water surface. The small cylinder has a higher melt rate than the large cylinder, which we can precisely predict from our theory since $U \propto R^{-1/4}$ (equation \ref{eq:cylmeltrate}) gives us a ratio of 1.24 in the melt rates which we also observe in our experiments within the experimental error (the value on the ordinate is just below \unit{10}{\micro\meter\per\second} for the $R_0 = \unit{25}{\milli\meter}$ cylinder and \unit{\approx 8}{\micro\meter\per\second} for the $R_0 = \unit{60}{\milli\meter}$ cylinder, giving a value of $\approx 1.2$, very close to our predicted value).

\subsection{ball}
\begin{figure}
    \centering
    \includegraphics[width=0.8\columnwidth]{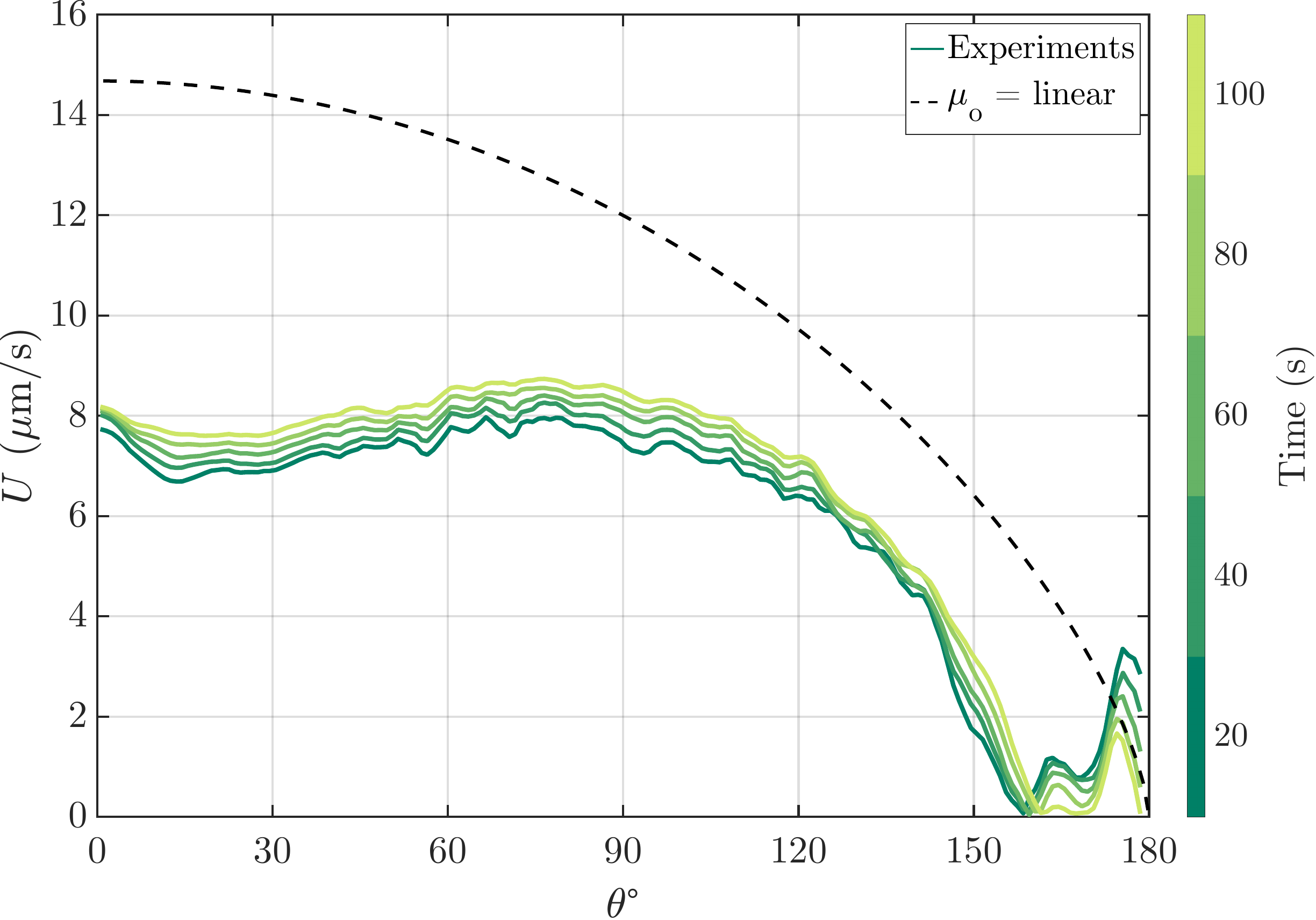}
    \caption{Cross-sectional inward melt-rate of a ball as a function of the angle $\theta$ for $\text{Ra}=\mathcal{O}(10^8)$. Note that $\theta = 0\degree$ is at the bottom of the ball. The solid green lines are results from the experiments. The black dashed line is the theoretical model for the spherical geometry. Note that now we only show the model with temperature-dependent viscosity. The reproducibility of this type of experiment is shown in the appendix. The video of this experiment can be found in the supplemental materials.}
    \label{fig:spheremodel}
\end{figure}

Finally we look at the melting of a ball with initial radius $R=\unit{60}{\milli\meter}$, see figure \ref{fig:contourcomp}b for the experiment and figure \ref{fig:theorycylinder} for the definition of the axes and other measurements. Figure \ref{fig:spheremodel} shows the melt rate for this ball and compares the experiments with the theory. Again, we restrict ourselves to the early stages of melting, when the assumption of a circular cross--section is still valid. Note that we also included temperature-dependent viscosity in our model. Like for the cylinders, the model shows reasonable agreement in terms of scaling and shape. Deviations for $\theta \leq 75\degree$ may be caused by the ambient water which we will discuss in section \ref{sec:discussion}. For high angles ($\theta \geq 160\degree$) the oil layer is much thicker and the flow is influenced by periodically-detaching droplets. In the dissolution based problem discussed in  \cite{DaviesWykes2018} the interface evolution differs greatly from what we show here in figure \ref{fig:contourcomp}. Their sugar balls dissolve into water, creating a Rayleigh--Taylor type instability similar to us. In our present work the instability is delayed due to the thin film of very viscous olive oil (as compared to water), the surface tension, and the curved surface, all modifying the instability. It is still unstable but with different wavelengths and growth rates, see e.g.\  \cite{gennes2004capillarity,trinh2014curvature,balestra2016rayleigh}. From the dispersion relation for semi-infinite bodies from e.g. \cite{Mikaelian1996} we get positive growth rate only for length scales from \unit{5}{\centi\meter} and larger. While our objects are bigger than \unit{\approx5}{\centi\meter}, it does not have sufficient time to grow to be apparent. We strictly see the pinch-off at the apex of our objects. For very large objects that are submerged in baths with smaller temperature differences (slow flow rates) this effect might be important.

\section{Theory} \label{sec:theory}
In this section we will derive the analytical models for the melt rate of the frozen olive oil objects that were shown in figures \ref{fig:wallmumodel}, \ref{fig:cylindermeltrates}, and \ref{fig:spheremodel}. We start with the theory for the vertical wall. After deriving this model we realized, from comparison with the experimental results, that it was needed to include temperature-dependent viscosity, and we incorporate this in the analysis. Using this as a starting point we then derive the theory for the cylindrical geometry and the spherical geometry.

\subsection*{Vertical wall}
\begin{figure}
    \centering
    \includegraphics[width=.5\columnwidth]{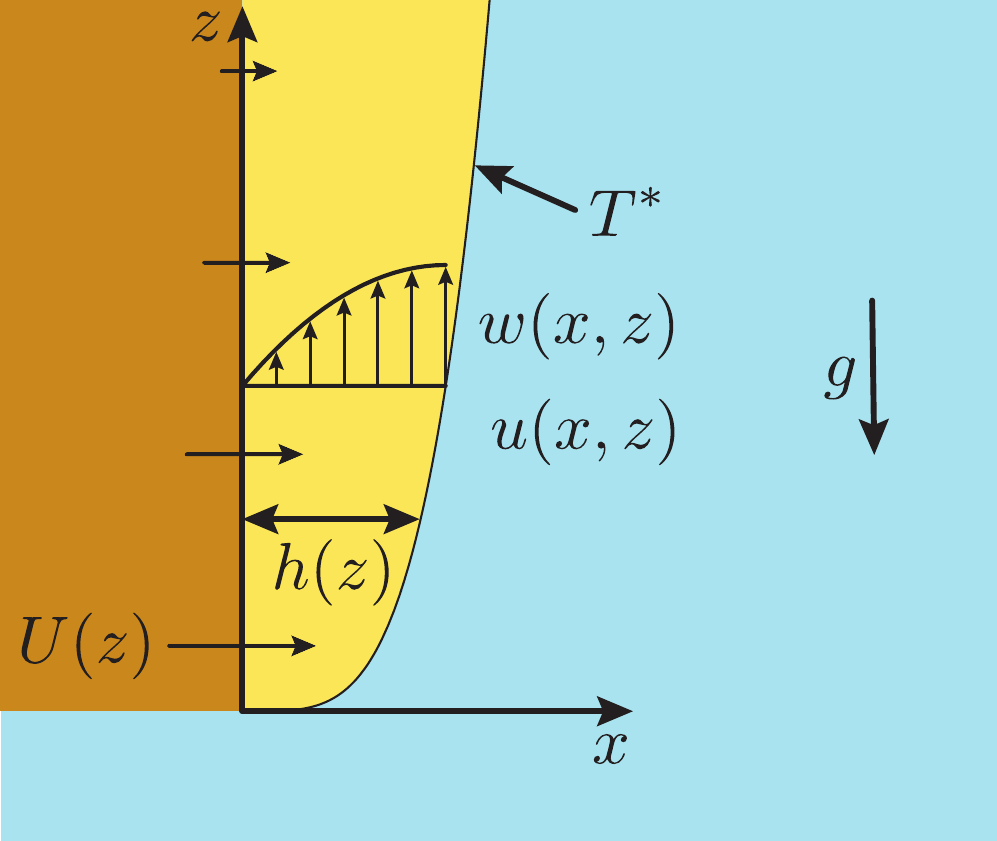}
    \caption{Schematic overview of melted olive oil (yellow) at a flat plate of frozen olive oil (orange), submerged in water. $U(z)$ is the melt rate at the frozen/liquid oil interface.}
    \label{fig:theorywall}
\end{figure}

\begin{figure}
    \centering
    \includegraphics[width=.5\columnwidth]{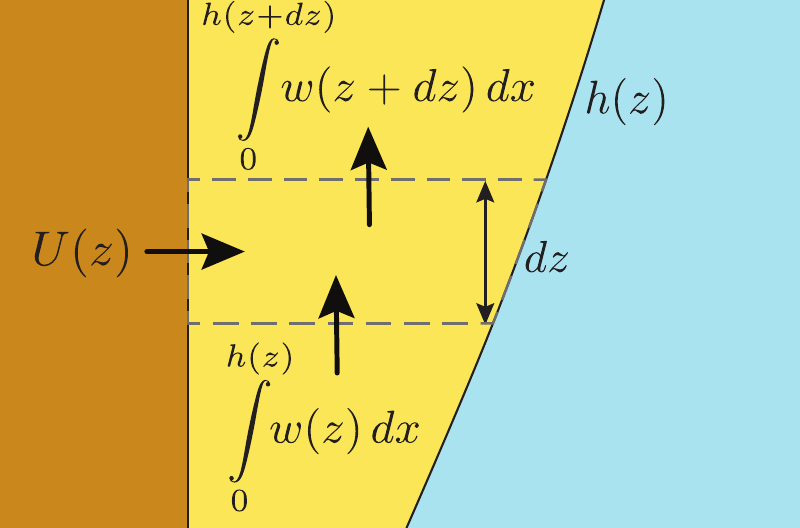}
    \caption{Illustration of the mass flows in a small control volume (dashed grey) of melting olive oil at a vertical wall. Oil is `injected' into the volume from the wall since the left of the control volume is following the interface, and oil enters the volume from below due to buoyancy forces. The total ingress from these two contributions equals the egress at the top of the control volume.}
    \label{fig:theoryslice}
\end{figure}

We start with the melting of the vertical wall, as shown in figure \ref{fig:theorywall}. With horizontal and vertical axes $x$ and $z$, respectively, the frozen wall is along the $z$-axis at $x=0$, the liquid melt layer is between $0 \leq x \leq h(z)$ while the ambient water, causing the oil to melt, stretches from $x=h(z)$ to infinity in $x$-direction. The surrounding water is cooling down as the oil is melting and therefore flows downward under the influence of gravity, reaching velocities in the order of $\unit{\!\!}{\centi\meter\per\second}$. These velocities are relatively low and as such we will assume that the ambient water is stationary. Inside the melt layer we have the horizontal velocity $u$ and the vertical velocity $w$ in the $x$ and $z$ directions, respectively. Gravity has acceleration $g$ in the negative $z$ direction. The most important are the densities $\rho_o$ and $\rho_w$ and the dynamic viscosities $\mu_o$ and $\mu_w$. Under these circumstances, where the oil is very viscous, and assuming we are in a steady state, the $w$-component of the Navier--Stokes equations simplifies such that we have a balance between the pressure gradient due to buoyancy and the viscous forces:
\begin{align}
    \mu_o \left(\frac{\partial^2w}{\partial x^2} + \frac{\partial^2w}{\partial z^2}\right) - g(\rho_o - \rho_w) = 0\label{eq:vertvel}.
\end{align}
Here we ignore surface tension effects, which we show in the discussion section to be a valid assumption. At the interface between solid and liquid oil we have $w(x=0) = 0$. The film has a thickness of $\mathcal{O}(\unit{\!\!}{\milli\meter})$, very small with respect to the height (\unit{30}{\centi\meter}), such that we are in the thin film limit ($\frac{\partial}{\partial x} \gg \frac{\partial}{\partial z}$) and we can therefore neglect the derivatives in the $z$-direction in equation \ref{eq:vertvel}. At the oil-water interface the stresses have to be balanced across the interface:
\begin{align}
    \mu_o \left. \frac{\partial w}{\partial x} \right|_{x=h^-} &= \mu_w \left. \frac{\partial w}{\partial x} \right|_{x=h^+},
\end{align}
where $h^-$($h^+$) indicates the gradient at the interface on the oil(water) side of the oil-water interface. Since our oil is much more viscous as compared to our water (table \ref{tab:liqprop}: $\mu_o/\mu_w = \mathcal{O}(100)$) and there is no reason to expect any sharp gradients in the water, we have for the velocity gradient at the oil-water interface:
\begin{align}
    \left. \frac{\partial w}{\partial x} \right|_{x=h^-} &= \frac{\mu_w}{\mu_o } \left. \frac{\partial w}{\partial x} \right|_{x=h^+} \approx 0. 
\end{align}
The solution of a simplified equation \ref{eq:vertvel} and obeying these boundary conditions then is:
\begin{align}
    w(x,z) = \beta x(2h(z)-x) \quad \text{ with: } \beta = \frac{g (\rho_w - \rho_o)}{2\mu_o} = \frac{g\Delta\rho}{2\mu_o}. \label{eq:solutionw1}
\end{align}
\noindent From the continuity equation follows the horizontal velocity:
\begin{align}
    \frac{\partial u}{\partial x} = -\frac{\partial w}{\partial z} = -2\beta x\frac{dh}{dz}.
\end{align}
We can now find the melt rate $U(z)$ at which the wall melts by integrating the previous equation to obtain:
\begin{align}
    u(x,z) = U(z) - \beta x^2\frac{dh}{dz}. \label{eq:horvel}
\end{align}
From mass conservation, see figure \ref{fig:theoryslice}, we can relate the melt rate and the film thickness:
\begin{align}
    U(z) = \frac{d}{d z}\left(\int_0^h wdx\right) = \frac{d}{dz}\left(\frac{2}{3}\beta h(z)^3\right).\label{eq:U(h)}
\end{align}
\noindent To find $h(z)$ we need to consider the thermal transport. The ambient water, with a temperature of $T_\infty = \unit{20}{\celsius}$ far away, transfers heat to the melt layer. This, in turn, transfers heat to the solid oil, causing this to melt further. We first consider the advection-conduction equation inside the thin melt layer. Assuming stationarity and using the thin film approximation we arrive at:
\begin{align}
    u\frac{\partial T}{\partial x} + w \frac{\partial T}{\partial z}= \alpha\frac{\partial^2T}{\partial x^2}, \label{eq:advdiff}
\end{align}
where $\alpha = (\lambda/\rho c_p)_o$ is the thermal diffusivity, $\lambda$ is the thermal conductivity, and $c_p$ is the specific heat capacity. In order to make the following analysis tractable\footnote{The Mises transformation (\cite{schlichting2016boundary}, page 183), where instead of $x$ and $z$ the coordinates $z$ and the stream function $\psi$ are used, leads to $\frac{\partial T}{\partial z} = \alpha \frac{\partial}{\partial \psi} \left( w \frac{\partial T}{\partial \psi} \right)$, which is no more tractable.}, we focus on the first term and neglect the second term of the left hand side of equation \ref{eq:advdiff}. For the first term we approximate $u$ by only taking the first term of equation \ref{eq:horvel}.
\begin{align}
    U(z)\frac{\partial T}{\partial x} = \alpha\frac{\partial^2T}{\partial x^2}\label{eq:adv-diff}.
\end{align}
After the result for $T(x,z)$ is obtained we will estimate the involved error in these approximations. One boundary condition is that $T$ equals the melting temperature $T_o$ at $x = 0$. At the interface of the melt layer and the ambient water, $x=h$, temperature and heat flux must be continuous. Since the velocity is small in both the oil and water, heat conduction is prominent. At the boundary, the water flowing down is only in contact with the wall for a short time (see the discussions in section \ref{sec:discussion}). Therefore, we approximate here the temperature at the interface between water and oil with the so-called contact temperature, occurring when two semi-infinite media with different temperature are brought in contact. With material properties $f = \sqrt{\lambda \rho c_p}$, this contact temperature is then (see e.g.~section 5.7, equation (5.63) from \cite{incropera1990fundamentals}):
\begin{align}
    T^* &= \frac{f_o T_o + f_w T_\infty}{f_o + f_w},  \label{eq:contacttemp}
\end{align}
where $T_\infty$ is the temperature of the water far away. Filling in the values for water and oil we get:
\begin{align*}
    T^* &= T_o + 0.75(T_\infty - T_o).
\end{align*}
The solution of equation \ref{eq:adv-diff} with $T_o$ at $x = 0$ and $T= T^*$ at $x=h$ results in
\begin{align}
    T(x,z) &= T_o + 0.75 (T_\infty - T_o) \frac{\int \limits_0^x e^{\frac{U(z)x}\alpha}dx}{\int \limits_0^h e^{\frac{U(z)x}\alpha}dx}.\label{eq:Tsol}
\end{align}
The heat flux at the wall, $\lambda(dT/dx)_{x=0}$, results in the melting rate. With latent heat $\mathcal{L}$ this means that
\begin{align}
    \lambda(dT/dx)_{x=0} = U(z)\rho_o\mathcal{L}\label{eq:fluxbalance}.
\end{align}
Combining equations \ref{eq:Tsol} and \ref{eq:fluxbalance} gives:
\begin{align}
    \frac{0.75\lambda(T_\infty - T_o)}{\int \limits_0^h e^{\frac{U(z)x}\alpha}dx} &= U(z) \rho_o \mathcal{L}. \label{eq:eleventwelve}
\end{align}
We now rewrite the expression and introduce the nondimensionalisation $\xi = x/h$:
\begin{align}
    0.75\lambda(T_\infty - T_o) &= U(z) \rho_o \mathcal{L} h \int \limits_0^1 e^{\frac{U(z)h \xi}\alpha}d\xi. \label{eq:huztemp}
\end{align}
Working out the integral and rearranging:
\begin{align}
    0.75\lambda(T_\infty - T_o) &= \rho_o \mathcal{L} \alpha \left[ e^{\frac{U(z)h}\alpha} -1 \right], \\
    \alpha \ln\left( 1+ \frac{0.75\lambda(T_\infty - T_o)}{\rho_o \mathcal{L} \alpha} \right) &= U(z) h = \frac{\beta}{2}\frac{d}{dz}\left(h^4\right), \label{eq:uzh}
\end{align}
where in the last step we used equation \ref{eq:U(h)}. We can now solve equation \ref{eq:uzh} for $h$ without further approximation:
\begin{align}
h(z) &= z^{1/4} \left[ \frac{2\alpha}{\beta} \ln(1+\Lambda) \right]^{1/4}, \label{eq:foundh}\\
    \Lambda &= 0.75(T_\infty - T_o)\frac{c_p}{\mathcal{L}} = 0.75\text{Ste} \label{eq:deflambda}.
\end{align}
We note that the dependence of the bath temperature on the layer thickness $h$ is very weak, not only is there a logarithmic dependence, but also the fourth root would experimentally verifying the dependence of $T_\infty$ on $h$ very challenging. Now that we have found an expression for $h$, $T$, and $U$ we can consider the approximations we have made before. We first focus on neglecting the term $w\frac{\partial T}{\partial z}$ in equation \ref{eq:advdiff}. Since $Uh/\alpha$ is in all our experiments small with respect to unity we may, from equation \ref{eq:Tsol}, for an estimate take $T-T_o$ as $C x/h$ with $C$ a constant. The main term, from equation \ref{eq:horvel}, $U(z)=2\beta h^2 \frac{dh}{dz}$. Hence the term $u\frac{\partial T}{\partial x}$ connected herewith is $2C\beta h \frac{dh}{dz}$. The term $w\frac{\partial T}{\partial z}$ is smaller, using equation \ref{eq:solutionw1} by an amount $\xi^2(2-\xi)/2$. This is significantly, though not negligibly, smaller; the error is $18.75\%$ at $\xi=1/2$. To obtain an estimate of the temperature distribution this is acceptable. We next consider the approximation of $u$ in equation \ref{eq:horvel}. We now use in equation \ref{eq:Tsol}, instead of $U$, the full expression in equation \ref{eq:horvel} and carry out the integration of the, then obtained version of equation \ref{eq:eleventwelve}, numerically, inserting the solution (equation \ref{eq:foundh}) for $h$. We find that the value of the integral then differs only $0.6\%$, owing to the fact that the argument of the exponent is $\ll 1$ for both cases (with and without the second term in equation \ref{eq:horvel}), such that the exponent is roughly constant $\approx 1$.

Finally we consider the neglect of convection in the transfer of heat between the oil film and water. In this connection we can consider the melt oil film as a solid since the velocity, as figures 3 and 5 show, is of order of \unit{10}{\micro\meter\per\second}, which is negligibly small. Water is cooled by the cold wall and flows downward. We can make an estimate of the velocity and the associated heat transfer by making use of the data on the opposite case, a hot wall and a cold fluid, since that is extensively dealt with in the literature. Following \cite{Bejan1993} (pages 345--346) for the case of the vertical wall, we define the Rayleigh number by $\text{Ra}_w = \epsilon_w g \left( T_\infty - T^* \right) \left(\frac L2\right)^3 \text{Pr}_w/\nu_w^2$, where, in addition to the quantities in table 1, $\epsilon=\unit{1.7\times 10^{-4}}{\kelvin^{-1}}$ is the thermal expansion coefficient of water at \unit{16.5}{\celsius}, $\nu_w=\mu_w/\rho_w$ the kinematic viscosity of water, and $\text{Pr}=\nu/\alpha$ the Prandtl number of water. $\text{Ra}_w$ depends on the location along the wall, for which we have taken halfway the height $L/2=\unit{0.15}{\meter}$. Using these quantities and table 1, we find $\text{Ra}_w=2.5\times 10^8$. With equation (7.43) from \cite{Bejan1993} (page 345) and $G$, from figure 7.5 (ibid. page 346) the downward velocity is \unit{1.04}{\centi\meter\per\second}. On this calculation the estimate of a water velocity of order of \unit{}{\centi\meter\per\second}, made earlier in this paper on page 8, was based. The corresponding Reynolds number ($\text{Re}=vL/\nu = 2800$) is far below $\mathcal{O}(10^5)$ where the flow becomes turbulent. \cite{Bejan1993}  (equation 7.51) gives for the above mentioned $\text{Ra}_w$ a Nusselt number $\text{Nu}=KL/(2\lambda_w)=58$, where $K$ is a heat transfer coefficient, with a value of about \unit{225}{\watt\per(\meter^2\kelvin)}. The associated heat flux from the oil film to the water is then \unit{1.6}{\kilo\watt\per\meter^2}. This is small with respect to the heat flux by conduction which is $\lambda_o (T^*-T_o)/h$. With $h\approx \unit{0.3}{\milli\meter}$ from (equation 4.17) this is \unit{\approx 12}{\kilo\watt\per\meter^2}. An additional and strong argument for the neglect of free convection is the following: consider equation \ref{eq:uzh}. The quantity $0.75(T_\infty – T_o)$ is $T^* - T_o$. In our theory this is a constant. If free convection where important $T^*$ would depend on $z$, and, as equation \ref{eq:uzh} shows, $h$ would not vary as $z^{1/4}$, but then deviate from that.

For comparison with experiments, $U(z)$ is the best quantity. From equations \ref{eq:U(h)} (or, alternatively, \ref{eq:uzh}) and \ref{eq:foundh} we have:
\begin{align}
    U(z) &= \frac{\alpha}{h}\ln{(1+\Lambda)} \propto z^{-1/4}, \label{eq:foundU}
\end{align}
which scales with the height $z$ to a power of $-1/4$. This exponent has been shown before in similar configurations by \cite{Wagner1949,ostrach1953analysis,Merk3,Wells2011}. We note that the melt rate only has a weak dependence on the bath temperature $T_\infty$. Combining equations \ref{eq:foundh} and \ref{eq:foundU}, we see that $U$ scales as the logarithm of $\Lambda$ taken to the power $3/4$. To achieve a doubling of our $U$ therefore means increasing our bath temperature to \unit{\approx 71}{\celsius}, which would create all kinds of problems with spurious flows created by heat leak with the surroundings causing natural convection.
The predicted melt rate in equation \ref{eq:foundU} is drawn in figure \ref{fig:wallmumodel} as the grey dotted line. This shows that the slope agrees satisfactory with the experimental data, but the values are considerably higher. We have now estimated the effect of three approximations viz.\ the reduction of $u$ to $U$, the neglect of $w\partial T/\partial z$ in equation 4.8 and the neglect of free convection at the water side of the melt film. None of these seems dominant as the source of the discrepancy between the theory and the measurements. Initially we suspected the neglect of the vertical heat transport to be the main cause. We made another estimate of the involved error by first writing equation 4.8 as:
\begin{align}
    \frac{\partial}{\partial x}\left(u T \right) + \frac{\partial}{\partial z}\left(w T \right) &= \alpha \frac{\partial^2 T}{\partial^2 x}
\end{align}
and subsequently integrating this term by term over the melt layer, using the solution of equation 4.11 for the temperature and $U$ and equation 4.4 for $u$ and $w$, respectively. Whereas the first term at the left and the conduction term both give $U(T^*-T_o)$ the second term on the left, the neglected term, results in $\frac U8 \left(T^* + 3T_o \right)$. With $T^*=\unit{13}{\celsius}$ and $T_o=\unit{-8}{\celsius}$, this term has a value of \unit{-11/8}{\kelvin\meter/\second}, in absolute magnitude only 6\% of $U(T^*-T_o)$. This again does not mark the neglect of vertical heat transport as causing the discrepancy. 

So far we had approximated the viscosity of the oil to be constant inside the melt layer, however, the strong temperature gradient inside the melt layer does not allow us to model the viscosity as constant. The temperature dependence of the viscosity of the oil is shown in figure \ref{fig:oilvisco} for the relevant temperature range. The temperature at the wall is \unit{-8}{\celsius} and the temperature at the oil-water interface is $T^* = \unit{13}{\celsius}$. The viscosity varies from \unit{380}{\centi P} to \unit{100}{\centi P} over this interval. In the following section we calculate again the vertical velocity $w$ in the melt, however, now taking a variable viscosity into account.

\subsection{Variation in viscosity} \label{sec:varvisco}
\begin{figure}
    \centering
    \includegraphics[width=0.6\columnwidth]{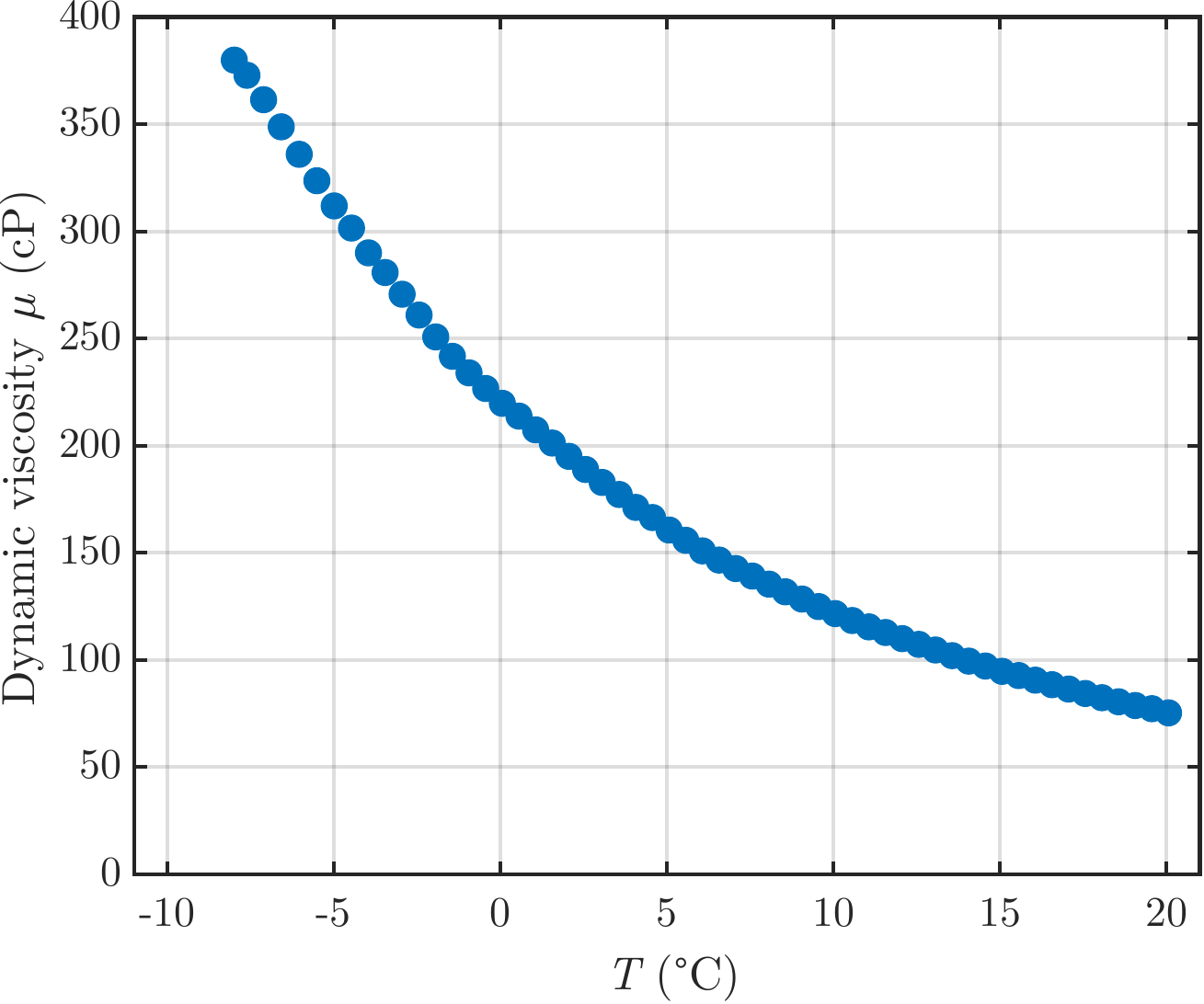}
    \caption{Olive oil viscosity as a function of the temperature. Measurements are done using an Anton Paar MCR502 rheometer with Peltier cooling. Measurements are done starting at a temperature of $\unit{20}{\celsius}$ and decreasing step-wise to a minimum of $\unit{-8}{\celsius}$. We estimate an error of $\unit{\pm 5}{\centi P}$.}
    \label{fig:oilvisco}
\end{figure}
In table \ref{tab:liqprop} we state a value for the viscosity of olive oil of \unit{170}{\centi P}. This value is taken at a mean olive oil temperature of \unit{3}{\celsius}. An Anton Paar MCR502 rheometer was used to measure temperature dependence of the viscosity of the olive oil, see figure \ref{fig:oilvisco}. As the oil is cooling down and approaching its freezing temperature its viscosity is increasing substantially. To account for the variation in the viscosity, we recalculate the vertical velocity $w$. In order to do this, we have to consider the Navier--Stokes equations where constant viscosity is not assumed and from that we obtain (different from equation \ref{eq:vertvel}):
\begin{align}
    \frac{\partial}{\partial x}\left(\mu_o(x)\frac{\partial w}{\partial x}  \right) = -g\Delta\rho.
\end{align}
Crucially here, we see that the viscosity is now inside the first derivative. After integrating once with respect to $x$ and rearranging we obtain:
\begin{align}
    \frac{\partial w}{\partial x} = \frac{-g\Delta\rho x}{\mu_o(x)} + C.\label{eq:mu1}
\end{align}
The boundary conditions remain unchanged:
\begin{align}
    w|_{x=0} &= 0, \label{eq:BC1} \\
    \frac{\partial w}{\partial x}\Bigg|_{x=h} &= 0. \label{eq:BC2}
\end{align}
We find the integration constant $C$ from using the boundary equation \ref{eq:BC2}:
\begin{align}
    \frac{\partial w}{\partial x} = \frac{g\Delta\rho (h-x)}{\mu_o(x)}. \label{eq:dwdz}
\end{align}
This can not be further solved analytically without assuming any particular shape for $\mu_o(x)$. We will follow two approaches. For the first, we will assume that the viscosity varies linearly inside the thin film to find the velocity profile $w$ which can be done fully analytically. For the second approach we will assume a linear \textit{temperature} profile inside the thin film, and by numerically integrating (including the viscosity curve of figure \ref{fig:oilvisco}) we find the velocity profile. 

For the first approach we will assume that the \textit{viscosity} is varying linearly inside the thin film:
\begin{align}
    \mu_o(x) = \mu_o(T_o)\left(1-\frac{x}{h}\right) + \mu_o(T^*)\frac{x}{h}.
\end{align}
We introduce the dimensionless quantity $\tilde{\mu}$:
\begin{align}
    \Tilde{\mu} &= \frac{\mu_o(T^*) - \mu_o(T_o)}{\mu_o(T_o)}.
\end{align}
Using this definition, equation \ref{eq:dwdz} can be rewritten and we obtain:
\begin{align}
    \frac{\partial w}{\partial x} &= \frac{2 \beta^*  h (h-x)}{h+\tilde{\mu} x}.
\end{align}
where $\beta^*$ is similar to $\beta$ (equation \ref{eq:solutionw1}) but with $\mu_o = \mu_o(T_o)$. After integrating and applying the boundary condition \ref{eq:BC1} we get:
\begin{align}
    w(x,z) &= 2 \beta^*  h^2 \left(\frac{\tilde{\mu}+1}{\tilde{\mu}^2} \ln \left(\frac{\tilde{\mu} x}{h}+1\right)-\frac{x}{h \tilde{\mu}}\right). \label{eq:wlinearvisco}
\end{align}
We see by expansion in $\tilde\mu$ that for $\tilde\mu = 0$, $w$ is the same as in equation \ref{eq:solutionw1}. Note that $w$ is still a function of both $x$ and $z$, since $h$ is a function of the height. As before, to get the melt rate $U(z)$ we integrate over a control volume in the liquid melt layer (see equation \ref{eq:U(h)}). 
\begin{align}
    U(z) &= \frac{d}{d z}\left(\beta^* \frac{2}{3}h(z)^3\right)\left(\frac{3 (1+\tilde{\mu})^2}{\tilde{\mu}^3} \ln (1+\tilde{\mu}) -\frac{6 + 9 \tilde{\mu}}{2 \tilde{\mu}^2}\right) \label{eq:linearmuresu}
\end{align}
Comparing this result with the previously found expression for melt rate $U(z)$ (r.h.s. of equation \ref{eq:U(h)}), it is seen that the varying viscosity introduces a correction on the melt rate that is dependent on the values for the viscosity at the wall and at the oil-water interface. Using equation \ref{eq:uzh} we see that the ratio between the melt rates for the constant viscosity versus the linearly varying viscosity is $\beta^*/\beta$ times the fourth root of the last bracketed term in equation \ref{eq:linearmuresu}. 

\begin{figure}
    \centering
    \includegraphics{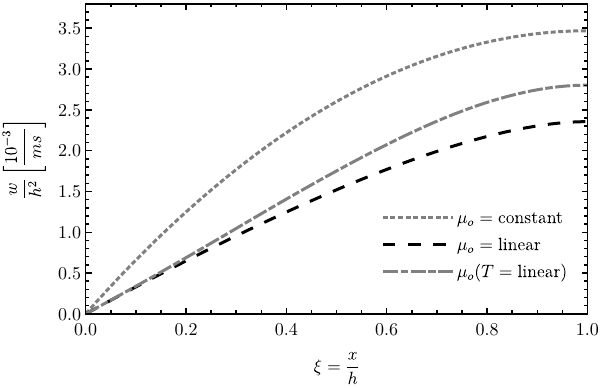}
    \caption{Vertical velocity profiles assuming constant viscosity (gray dotted, equation \ref{eq:solutionw1}), linear viscosity (black dashed, equation \ref{eq:wlinearvisco}), and linear temperature profile and viscosity following figure \ref{fig:oilvisco} (gray dash-dotted, equation \ref{eq:fullw}). }
    \label{fig:wprofiles}
\end{figure}

For the second approach we will assume that the \textit{temperature} is varying linearly inside the thin film:
\begin{align}
    T(\xi) &= \xi T^* + (1-\xi) T_o
\end{align}
We can now fill this into equation \ref{eq:dwdz} and integrate to find $w$:
\begin{align}
    \frac{\partial w}{\partial x} &= \frac{g\Delta\rho (h-x)}{\mu_o(T(\xi))}, \\
    \frac{w(\xi)}{h^2} &= g\Delta\rho \int \frac{1-\xi}{\mu_o(T(\xi))} d\xi + C. \label{eq:fullw}
\end{align}
The curve $\mu_o(T)$ from figure \ref{fig:oilvisco} is used to numerically integrate the profile and using the boundary condition (equation \ref{eq:BC1}) the integration constant can be determined. The comparison of all these velocity profiles can be found in figure \ref{fig:wprofiles}. Going from a constant viscosity to a viscosity that varies linearly dramatically decreases the velocity $w$ and stays relatively close to a parabolic profile. If we now vary the viscosity according to the curve of figure \ref{fig:oilvisco} (on average this viscosity is lower than the linear approximation) the curve is slightly higher but the shape of the profile is still roughly parabolic for the viscosities under consideration.

We can now find $U$ for the general case of $\mu(T(\xi))$ by rewriting equation \ref{eq:U(h)}:
\begin{align}
    U(z) &= \frac{\partial}{\partial z}\left(h^3\int_0^1 \frac{w}{h^2}d\xi\right) = \frac{\partial}{\partial z}\left(h^3 \right) \int_0^1 \frac{w}{h^2}d\xi, \label{eq:uzgeneral}
\end{align}
where we note that the integral is only a function of $\xi$ (see equation \ref{eq:fullw}) and evaluates to a constant, and can therefore be taken outside the derivative. We evaluate now $hU(z)$:
\begin{align}
    h U(z) &= \frac 34 \frac{\partial}{\partial z}\left(h^4 \right) \int_0^1 \frac{w}{h^2}d\xi,
\end{align}
which we can now insert into equation \ref{eq:huztemp} to find an equation similar to equation \ref{eq:uzh}:
\begin{align}
       \alpha \ln\left( 1+ \frac{0.75\lambda(T_\infty - T_o)}{\rho_o \mathcal{L} \alpha} \right) &= U(z) h = \frac 34 \frac{\partial}{\partial z}\left(h^4 \right) \int_0^1 \frac{w}{h^2}d\xi.
\end{align}
We can solve for $h$ (again realising that the last integral evaluates to a number and does not depend on $z$ nor $h$):
\begin{align}
       h(z) &= z^{1/4} \left[ \frac{\alpha\ln(1+\Lambda)}{\frac 34 \int_0^1 \frac{w}{h^2}d\xi} \right]^{1/4}
\end{align}
We now plug this in equation \ref{eq:uzgeneral} to obtain:
\begin{align}
     U(z) &= \frac{\partial}{\partial z} z^{3/4} \left[ \frac{\alpha\ln(1+\Lambda)}{\frac 34 \int_0^1 \frac{w}{h^2}d\xi} \right]^{3/4} \int_0^1 \frac{w}{h^2}d\xi \\
     U(z) &= \frac{1}{z^{1/4}} \left[\alpha\ln(1+\Lambda) \right]^{3/4} \left[\frac 34 \int_0^1 \frac{w}{h^2}d\xi \right]^{1/4} \label{eq:uzfinalresult}
\end{align}
One can see that equation \ref{eq:uzfinalresult} simplifies to equation \ref{eq:foundU} once we take $\mu_o$ constant which allows us to integrate equation \ref{eq:fullw} analytically (to obtain a half-parabolic profile), and integrating that once more with respect to $\xi$ to obtain the vertical flux.

\subsection*{Cylinder}
\begin{figure}
    \centering
    \includegraphics[width=.5\columnwidth]{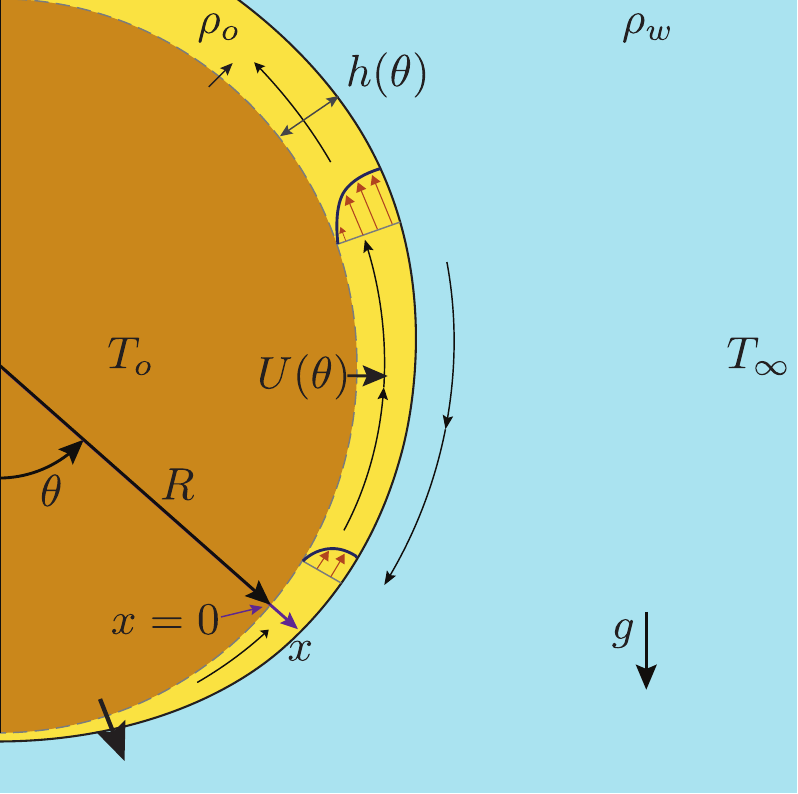}
    \caption{Schematic overview of molten olive oil (yellow) over a circular shape of frozen olive oil (orange), submerged in water (blue). $U(\theta)$ is the melt rate at the frozen/liquid oil interface. Definitions are used for the horizontal cylinder (polar coordinates) and for the ball (spherical coordinates).}
    \label{fig:theorycylinder}
\end{figure}

For the horizontal cylinder (see figure \ref{fig:contourcomp}a), we need to perform a coordinate transform to polar coordinates, see figure \ref{fig:theorycylinder}. The buoyancy term $\beta$ now depends on the angle $\theta$ with the vertical direction, and the coordinate transform is applied to the governing equations. The melt rate $U$ and film thickness $h$ now depend on the angle $\theta$ instead of height $z$. Since $h \ll R$ with $R$ the radius of the cylinder, the tangential velocity (with $x = r-R$) has the same profile as $w$ for the vertical wall. Equation \ref{eq:contrad} shows the continuity equation in polar coordinates where the axial dependence has been assumed absent, where $u_r$ is the radial velocity, and $u_\theta$ is the tangential velocity, with $u_\theta = \bar{\beta}x(2h(\theta)-x)$ analogous to the vertical velocity $w$ for the vertical wall, where $\mu_o$ is taken constant conform equation \ref{eq:solutionw1}. $\bar{\beta} = \beta\sin{\theta}$ is the buoyancy term adapted to the geometry.

\begin{align}
    \frac{\partial}{\partial x}\big((R+x)u_r \big) = \frac{\partial}{\partial\theta}(u_\theta). \label{eq:contrad}
\end{align}
Substituting $u_\theta$ in equation \ref{eq:contrad} and using an analogous boundary condition to the vertical wall case, $u_r(x=0) = U(\theta)$, results in an expression for $u_r$:
\begin{align}
    u_r(\theta,x) = U(\theta) + \frac{1}{R}\left(-\frac{1}{3}\frac{d}{d\theta}(\bar\beta)x^3 +\frac{d}{d\theta}(\bar\beta h)x^2 \right)\label{eq:urlong}.
\end{align}
An expression for the melt rate can be obtained by considering mass conservation:
\begin{align}
    U(\theta) = \frac{1}{R}\frac{\partial}{\partial\theta}\int_0^{h(\theta)}u_\theta(\theta,x)dx = \frac{1}{R}\frac{d}{d\theta}\left(\frac{2}{3}\bar\beta h^3\right)\label{eq:meltsphere}.
\end{align}
The advection-conduction equation, analogous to equation \ref{eq:adv-diff}, with $\xi = \frac{x}{h}$, follows:
\begin{align}
    u_r\frac{\partial T}{\partial \xi} = \frac{\alpha}{h}\frac{\partial^2 T}{\partial \xi^2}.
\end{align}
For the temperature we make an approximation analogous to the case of the vertical wall, i.e.\ $u_r \approx U(\theta)$. The boundary conditions are analogous to the case of a vertical wall:
\begin{align}
    T|_{\xi=0} &= T_o,\\
    T|_{\xi=1} &= T_o + 0.75\Delta T
\end{align}
where $\Delta T = T_\infty - T_o$. The solution for the temperature profile is:
\begin{align}
    T(\xi) = T_o + 0.75\Delta T \frac{ \int_0^\xi \exp{ \left[ \frac{2h}{3R\alpha}  \frac{d}{d \theta} \left( \bar\beta h^3 \right) \xi  \right] d\xi } }{ \int_0^1 \exp{\left[ \frac{2h}{3R\alpha} \frac{d}{d \theta} \left(\bar\beta h^3\right)\xi  \right]d\xi}}\label{eq:Txi}.
\end{align}
Analogous to equation \ref{eq:fluxbalance} the heat flux balance is:
\begin{align}
    \rho_oU(\theta)\mathcal{L} = \frac{0.75\lambda \Delta T}{h} \frac{1}{\int_0^1\exp{\left[\frac{2 \bar\beta h}{3R\alpha}\frac{d}{d\theta}\left(\sin(\theta) h^3\right)\xi \right]} d\xi}.
    \label{eq:fluxsphere}
\end{align}
We define the quantities
\begin{align}
    q^4 = \frac{R\alpha}{\beta},\;\;\text{and}\;\; H = \frac{h^4}{q^4}\label{eq:q}.
\end{align}
Then the integral in equation \ref{eq:fluxsphere} becomes
\begin{align}
    \int_0^1 \exp{\left[ \left(\frac{1}{2}\frac{dH}{d\theta}\sin(\theta) + \frac{2}{3}H \cos(\theta)\right)\xi \right] }d\xi
    \label{eq:integrandsimple}.
\end{align}
We define a function $G(\theta)$:
\begin{align}
    G(\theta) = \frac{2}{3}H\cos(\theta) + \frac{1}{2}\frac{d H}{d \theta}\sin(\theta)\label{eq:G}.
\end{align}
Such that we can write the integral as:
\begin{align}
    \int_0^1\exp{\left[ G(\theta)\xi \right]}d\xi\label{eq:Gint}.
\end{align}
From equations \ref{eq:meltsphere}, \ref{eq:q}, and \ref{eq:G} we deduce that
\begin{align}
    hU(\theta) = \alpha G(\theta)\label{eq:UG}.
\end{align}
Inserting this into equation \ref{eq:fluxsphere} and using equation \ref{eq:Gint} results in
\begin{align}
    0.75 \text{Ste} = \exp(G(\theta)) - 1.
\end{align}
With equation \ref{eq:deflambda} and taking logarithms we arrive at:
\begin{align}
    G(\theta) = \frac{2}{3}H\cos(\theta) + \frac{1}{2}\frac{d H}{d \theta}\sin(\theta) = \ln{(1+\Lambda)}.
\end{align}
Solving and requesting regularity at $\theta = 0$ gives:
\begin{align}
    H = \left( \frac{h}{q}\right)^4 = 4\ln{(1+\Lambda)}\frac{ \int_0^\theta \sin(\theta^*)^{1/3}d\theta^*}{\sin(\theta)^{4/3}}.\label{eq:H}
\end{align}
The final solution for the melt film thickness and melt rate can now be found by rewriting equation \ref{eq:H}, using equation \ref{eq:q}, and substituting the result for the melt film thickness in equation \ref{eq:UG}:
\begin{align}
    h_{\text{cyl}}(\theta) &= R^{1/4}\left[ \frac{2\alpha}{\beta}\ln{(1+\Lambda)} \right]^{1/4}f_{\text{cyl}}(\theta) \label{eq:cylh}, \\
    U_{\text{cyl}}(\theta) &= \frac{\alpha}{h_{\text{cyl}}}\ln{(1+\Lambda)}. \label{eq:cylmeltrate}
\end{align}
where $f_{\text{cyl}}(\theta)$ is a shape function:
\begin{align}
    f_{\text{cyl}} &= \frac{\left( \int_0^\theta \sin(\theta^*)^{1/3}d\theta^*\right)^{1/4}}{\sin(\theta)^{1/3}}, \\
    &= \begin{cases}
        \frac{\sqrt[4]{3\;{}_2F_1\left(\frac{1}{2},\frac{2}{3};\frac{5}{3};\sin(\theta)^2\right)}}{\sqrt{2}} \;\;\;\; &0 \leq \theta \leq \frac{\pi}{2}\\
        \frac{\sqrt[4]{\frac{3 \sqrt{3} \Gamma \left(\frac{2}{3}\right)^3}{2^{2/3} \pi } -\frac{3}{4}\sin(\theta)^{\frac{4}{3}} {}_2F_1\left(\frac{1}{2},\frac{2}{3};\frac{5}{3};\sin(\theta)^2\right)}}{\sqrt[3]{\sin(\theta)}} \;\;\;\; &\frac{\pi}{2} \leq \theta \leq \pi 
    \end{cases},
\end{align}
where ${}_2F_1$ is Gauss's hypergeometric function and $\Gamma$ the complete gamma function. Compared to the expressions that were found for the vertical wall, the difference is in this shape function, which compensates for the geometry varying when following the boundary of the wall, and the dependence on the radius $h_\text{cyl} \propto R^{1/4}$. Similar expressions occur in \cite{Acrivos1960AFluids,Acrivos1960MassVelocities} and also resemble the solutions for a dissolving \textit{vertical} cylinder found by \cite{Pegler2020}.

\subsection*{ball}
The problem of a melting ball is very similar to the cylinder described above. $\theta = 0$ is again defined on the bottom side of the object, the azimuthal angle $\phi$ is defined positive in clockwise direction, and $x = r-R$, with $x=0$ at the surface, is defined in the same manner as the horizontal cylinder, see figure \ref{fig:theorycylinder}. An important difference is a flow focusing  due to the varying circumference of the ball with changing polar angle $\theta$. The continuity equation in spherical coordinates, where $u_\theta  = \bar\beta x(2h-x)$ (again assuming $\mu_o$ is constant), is unchanged and azimuthal symmetry is assumed:
\begin{align}
    \frac{1}{r^2}\frac{\partial}{\partial r}(r^2u_r) + \frac{1}{r\sin(\theta)}\frac{\partial}{\partial\theta}(u_\theta\sin(\theta)) = 0\label{eq:contsphere}.
\end{align}
From this, $u_r$ can be found as before:
\begin{align}
    u_r = U(\theta) - \bar \beta \left(2 h\cos(\theta) + \frac{d h}{d\theta}\sin(\theta)\right)\frac{x^2}{R} + \frac{2}{3}\bar\beta \cos(\theta)\frac{x^3}{R}.
\end{align}
The control volume over the film thickness is now taken in three dimensions:
\begin{align}
    2\pi R^2\sin(\theta) U(\theta) = \frac{d}{d\theta}\int_0^h2\pi R\sin(\theta)u_\theta dx,
\end{align}
which after substitution of $u_\theta$ gives:
\begin{align}
    U(\theta) = \frac{1}{R\sin(\theta)}\frac{\partial}{\partial \theta}\left(\frac{2}{3}\bar\beta h^3\sin(\theta)^2\right).
\end{align}
Following a similar procedure as before, we find a new function $G(\theta)$:
\begin{align}
    G(\theta) = \frac{4}{3}\cos(\theta) H + \frac{1}{2}\sin(\theta)\frac{dH}{d\theta} = \ln{(1+\Lambda)}.
\end{align}
This can be solved to obtain the final solutions for the film thickness and the melt rate:
\begin{align}
    h_{\text{ball}}(\theta) &= R^{1/4}\left[ \frac{2\alpha}{\bar\beta}\ln{(1 + \Lambda)} \right]^{1/4}f_{\text{ball}}, \label{eq:sphereh}\\
    U_{\text{ball}}(\theta) &= \ln{(1 + \Lambda)}\frac{\alpha}{h_{\text{ball}}}, \label{eq:spheremeltrate}
\end{align}
where $f_{\text{ball}}$ is a shape function for the spherical geometry, different from the one for the cylinder:
\begin{align}
    f_{\text{ball}} &= \frac{\left( \int_0^\theta \sin(\theta^*)^{5/3}d\theta^*\right)^{1/4}}{\sin(\theta)^{2/3}},\\
    &= \begin{cases}
        \frac{\sqrt[4]{3\;{}_2F_1\left(\frac{1}{2},\frac{4}{3};\frac{7}{3};\sin(\theta)^2\right)}}{2^{3/4}} \;\;\;\; &0 \leq \theta \leq \frac{\pi}{2}\\
        \frac{\sqrt[4]{\frac{\sqrt{3} \Gamma \left(\frac{1}{3}\right)^3}{5 \sqrt[3]{2} \pi } -\frac{3}{8}\sin(\theta)^{\frac{8}{3}} {}_2F_1\left(\frac{1}{2},\frac{4}{3};\frac{7}{3};\sin(\theta)^2\right) }}{\sin(\theta)^{\frac{2}{3}}} \;\;\;\; &\frac{\pi}{2} \leq \theta \leq \pi 
    \end{cases}.
\end{align}

We like to highlight that the solutions for the thickness $h$ (equations \ref{eq:foundh}, \ref{eq:cylh}, and \ref{eq:sphereh}) and for the melt rate $U$ (equations \ref{eq:foundU}, \ref{eq:cylmeltrate}, and \ref{eq:spheremeltrate}) for all three geometries have a very similar form. 

\section{Discussion} \label{sec:discussion}
In section 4 we already discussed some approximations that we made. We have shown that our models match relatively well with our experimental findings, especially considering that our model does not contain any free (fitting) parameters. During the derivation we made several approximations, and the model does not include all effects. We will now go through the remaining approximations and assess their validity.

First, we have made use of the thin film approximation, which seems like a reasonable approximation since our layer thickness is of $\mathcal{O}(\milli\meter)$ while our objects are of $\mathcal{O}(\unit{100}{\milli\meter})$. 

Second, we had made the assumption that $\left. \frac{\partial w}{\partial x} \right|_{x=h} \approx 0$ which also seems justified since the ratio of the viscosities, even in the worst case, is $\mu_o/\mu_w \approx 75 \gg 1$.

Third, the assumption of a constant contact temperature $T^*$ along the wall, turns out to be realistic for the vertical wall, and for the cylinder, as can be seen from figures \ref{fig:wallmumodel} and \ref{fig:cylindermeltrates}. In the case of the ball the agreement with the model is good between $\unit{60}{\degree}$ and $\unit{130}{\degree}$ but there is a significant difference in the bottom region, see figure \ref{fig:spheremodel}. The reason for that becomes clear from the following analysis. When two semi-infinite media of different temperatures are brought in contact the interface assumes a temperature, the contact temperature $T^*$, given in equation \ref{eq:contacttemp}, which remains constant thereafter. In our case one of the media, the melt film, is of finite extent $h$. If we take, for convenience, equal material properties at both sides, the contact temperature changes in time according to (see Appendix \ref{sec:finitehtstar})
\begin{align}
    T^* &= T_o + \frac{T_\infty - T_o}2 \erf\left(\frac h{\sqrt{\alpha t}}\right), \label{eq:erf}
\end{align}
where $\erf(\tau) = \frac 2{\sqrt{\pi}} \int_0^\tau e^{-t^2}dt$ is the error function. Given a typical time of $\unit{20}{\second}$ of contact of a water element from top to bottom of the ball, and a $h = \unit{2}{\milli\meter}$, with $\alpha = \unit{8\cdot 10^{-8}}{\meter^2\per\second}$, this means that after $\unit{20}{\second}$ the error function in equation \ref{eq:erf} has still 96\% of its initial value. However, near the bottom, where $\theta=0$ and $h$ is very small, of the order of tenths of millimeters, and where probably the contact time is longer, the error function decreases from its initial value. This means a drop in the contact temperature and thereby of the melting rate. For the vertical wall the film thickness is of the order of a $\unit{1}{\milli\meter}$ along most of the wall. Unfortunately we are unable to locally measure the temperature profiles since the scales are too small (and probes too big).

Four, throughout the analysis we had assumed that the problem is time-independent. Since our freezing temperature is $T_i = \unit{-14}{\celsius}$ and the melting point it $T_o = \unit{-8}{\celsius}$ all the matter has to warm up $\unit{6}{\kelvin}$ before it melts. At $t=0$ a skin layer of the temperature grows inside the material. The typical dimensionless similarity variable $\eta = x/\sqrt{\alpha t}$ can then be used to find the temperature profile inside the material which goes like $\erf(\eta/2)$. The typical penetration depth of the temperature is thus given by $\sqrt{\alpha t}$, such that the speed at which this front moves is $U_{\text{skin layer}} = \sqrt{\frac{\alpha}{4t}}$. If we equate this to our melt speed of $U \approx \unit{10}{\micro\meter\per\second}$ we find a typical time scale of $t \approx \unit{200}{\second}$ at which the speed of the penetrating skin layer reaches the same speed as the melting boundary. In other words, for times below a few minutes, there is energy spend on heating up material that is not melted in this time. For larger times, the speed of skin layer and the melting boundary moving along at the same speed, and energy is only spent on heating material that is also melted. This thus means that for small times we overpredict the melting rate, and the actual melting rate is slightly less since we heat more material up than we melt. Note that the energy spent on heating is relatively small as compared to the energy spent on the phase transition $c_p(T_i - T_o)/\mathcal{L} = 4.4\%$, such that the effect is comparatively small. This is, however, a clear case in which our prediction causes an over prediction of the melt rate. A small correction could be made by introducing an effective latent heat. The idea is to include not only the energy for the phase change (the latent heat), but also the energy needed to heat up the material from its initial temperature of $\unit{-14}{\celsius}$ to the melting point $\unit{-8}{\celsius}$. Heating up the oil by $\Delta T = \unit{-8}{\celsius} - (\unit{-14}{\celsius}) = \unit{6}\kelvin$ prior to melting requires an energy of $c_p\Delta T = \unit{11.82}{\kilo\joule\per\kilo\gram}$. Comparing this with the latent heat of olive oil $\unit{267}{\kilo\joule\per\kilo\gram}$, this amount to roughly 4.4\%. An effective latent heat could then be defined that takes into consideration: $\mathcal{L}_\text{eff} = \mathcal{L}+c_p(T_i - T_o)$.

Another experimental issue not yet discussed in detail, might be that, as we remove the frozen oil from the metallic mould and then place in our water tank the oil has slightly heated up. We are not sure whether all objects were $\unit{-14}{\celsius}$ throughout. Whereas the melting profiles $U$ are more or less constant for the vertical wall and the large horizontal cylinder, for the ball and the small cylinder the melting profiles $U$ change a bit over time. The reason between those could be the varying time between releasing the olive oil from the mould and placing them in our aquarium, see figure \ref{fig:expsetup}. We hypothesize that for the vertical wall and the large cylinder the object was left (relatively) long in air and would already start forming the temperature skin layer. For the experiments with the small cylinder and the ball we quickly used the object after releasing it from the mould, not allowing the temperature skin layer to develop. This could then explain the differences in steadiness between the melting profiles $U$ for the various geometries.

Fifth, up to now we have ignored surface tension effects. To justify this assumption, we compare the hydrostatic pressure and the Laplace pressure. Typical difference in hydrostatic pressures (for the vertical wall at mid-height) is given by: $P_\text{hydrostatic} = g\Delta\rho L/2 \approx \unit{200}{\pascal}$. Now the Laplace pressure can be approximated as $P_\text{Laplace} = \sigma \left|d^2h/dz^2 \right| = \unit{0.2}{\milli\pascal}$. Since the Laplace pressure is smaller by 6 orders of magnitude, we are confident in ignoring the effects of surface tension. This ratio is substantially less for the cylindrical and spherical experiments; $\approx 200$ and $\approx 500$ respectively, but still high enough to be confident in ignoring effects of surface tension. However, surface tension might be important for the spherical and cylindrical geometries for $\theta > \unit{90}{\degree}$ to stabilise the oil film.

Sixth, we observe no turbulence in the ambient fluid nor in the thin oil film, such that we expect our results to be valid at least up to $\text{Ra}=10^8$. Since our Stefan number is relatively low (0.21) the heating up of the oil is small as compared to the heat needed to melt. For high Stefan numbers we expect an additional conduction problem inside the objects that, for small $t$, delays the melting. So we can therefore conclude that our results should also be valid for Stefan numbers at least up to $\text{Ste}=0.2$.

Lastly, throughout our derivation we have considered the ambient water to be stationary, which we have discussed before. The velocity in the oil layer along the wall is of order \unit{1}{\milli\meter\per\second} and the velocity in water \unit{1}{\centi\meter\per\second}. This is so low that heat transfer is still mainly by conduction and will influence $T^*$ in a small amount. However, for applications with much larger Rayleigh and Stefan numbers this situation might change, and our analysis and assumptions will no longer hold. For example: increased velocities in the surrounding water may affect the pinch-off at the apex due to increased shear forces, or the attached oil film may show instabilities which might result in detachment of the oil film from the object prior to reaching the apex.

\section{Conclusion} \label{sec:conclusion}
In this work we have studied the melting process of frozen olive oil in an immiscible environment of water for Rayleigh numbers of order $\mathcal{O}(10^8)$ and Stefan number $\text{Ste} = 0.21$. We have studied three different geometries with different symmetries experimentally and model the melt rate along the interface. Our model can predict the height (or angular) dependence of the melt rate for the three geometries, and not only the scaling but also the prefactor can be reasonably predicted (correct order of magnitude). In our derivation of the model, we highlight the importance of the approximation to the viscosity and its dependence on temperature. As discussion in the prior section, none of the approximations made in the model appears to be the main cause of the discrepancy with the experiments. We do not observe the sharpening that had been observed by \cite{Huang2022} in the immiscible case, the pinching off at the tip (either for the case of a horizontal cylinder or a ball) hinders the smooth flow that would otherwise sharpen it.

\section{Acknowledgements}
We thank Gert-Wim Bruggert, Martin Bos, Dennis van Gils, and Thomas Zijlstra for technical support. We also acknowledge fruitful discussions with Detlef Lohse, Howard Stone, and Andrea Prosperetti. This work was financially supported by The Netherlands Center for Multiscale Catalytic Energy Conversion (MCEC), an NWO Gravitation Programme funded by the Ministry of Education, Culture and Science of the government of The Netherlands, and the European Union (ERC, MeltDyn, 101040254). The authors report no conflict of interest. 


\clearpage
\appendix
\section{Effect of finite $h$ on $T^*$} \label{sec:finitehtstar}
Consider a piece of material of length $h$, and temperature $T_o$, lying between $x=0$ and $x=h$. This is at time $t=0$ brought into contact with a semi-infinite piece of the same material, between $x=h$ and $x=\infty$, and at temperature $T_\infty$. The side of the first piece at $x=0$ is kept at $T_o$ at all times. We are interested in the temperature at $x=h$. With $\alpha$ as defined in the text, the heat equation in both pieces is
\begin{align}
    \frac{\partial T}{\partial t} &= \alpha \frac{\partial^2 T}{\partial x^2}.
\end{align}
Applying a Laplace transform $L(t) = \int_0^\infty T e^{-st} dt$, taking into account the above mentioned 
boundary and initial conditions gives for the solution of the Laplace transform of the temperature at $x=h$:
\begin{align}
    L(T(x=h)) &= \frac{T_\infty - T_o}{2} \frac{e^{-2h\sqrt{\frac{s}{\alpha}}}}s.
\end{align}
Using a table of inverse Laplace transforms results for the temperature at $x=h$ in
\begin{align}
    T^* &= T_o + \frac{T_\infty - T_o}2 \erf\left(\frac h{\sqrt{\alpha t}}\right).
\end{align}


\section{Reproducibility} \label{sec:reprod}

\begin{figure}
    \centering
    \includegraphics[width=.75\columnwidth]{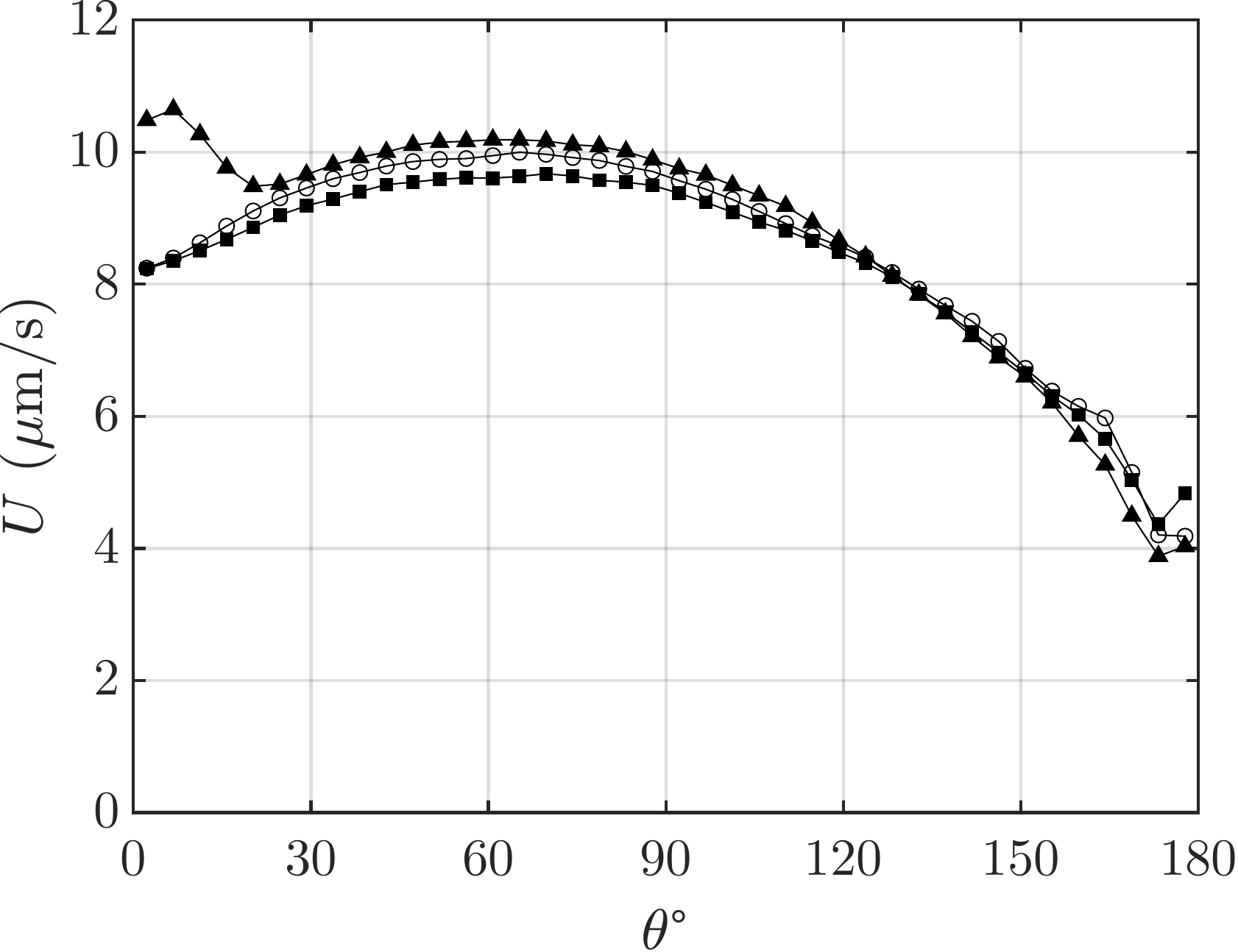}
    \caption{The average melting rate $U$ for $\unit{600}{\second} \leq t \leq \unit{960}{\second}$ as function of the angle $\theta$ for three comparable experiments of a melting ball with initial radius $R=\unit{60}{\milli\meter}$ corresponding to $\text{Ra}=\mathcal{O}(10^8)$.}
    \label{fig:reproducibilitysphere}
\end{figure}

We have done several experiments for each geometry with similar results, though the experimental conditions are generally slightly different (freezing temperature, water temperature, time outside the mould before being inserted in the aquarium, camera field of view, recording frame rate etc.). We have found three experiments with very comparable experimental conditions for the spherical geometry to show the reproducibility of the experiments. Though the recording frame rates are different their least common multiple is \unit{60}{\second}, such that we can compare frames at this interval. To release the balls from their moulds we have to slightly heat the mould to melt a very thin layer of oil. After that, it can take some time to screw the ball to the holder and to place the ball in the aquarium, and there can be slightly different fluid motion inside the tank due to the insertion of the oil ball in to the tank. Combining these effect gives a slight uncertainty in the exact starting temperature (though all close to \unit{-14}{\celsius}), and slightly different initial conditions in terms of flow. We therefore opt to look at the balls at a slightly later stage where these transient effect are less important. Whereas figure \ref{fig:spheremodel} shows multiple times for a single experiment, we now focus on the average melting rate over a single time interval, $\unit{600}{\second} \leq t \leq \unit{960}{\second}$, for several experiments to keep the figure legible, see figure \ref{fig:reproducibilitysphere}. We find reasonable reproducibility between the different experiments. For $\theta \geq \unit{160}{\degree}$ (top region) we see more spread between the experiments as there the droplets are pinching off. For $\theta \leq \unit{20}{\degree}$ we also see the triangle dataset to be different from the other two. We observe a slightly flattened bottom in the images which can explain this discrepancy, which is perhaps due to an air bubble or crack during the unmoulding and freezing process. Lastly, the water temperature is slightly higher for the triangle experiment ($T_\infty = \unit{19.2}{\celsius}$) as compared to the other two experiments ($T_\infty = \unit{18.5}{\celsius}$).

\end{document}